\documentstyle[preprint,amsfonts,aps]{revtex}
\begin{document}
\draft
\preprint{\vbox{\hbox{\large G\"oteborg ITP 94-20}
                \hbox{\large Submitted to Phys. Rev. B}}}

\title{\Large\bf Spin Dynamics of\\
                 the Triangular Heisenberg Antiferromagnet:\\
                 A Schwinger Boson Approach}

\author{\large Ann Mattsson}

\address{\normalsize\em Institute of Theoretical Physics\\
Chalmers University of Technology and\\
G\"oteborg University\\
S-412 96 G\"oteborg, Sweden}

\maketitle
\begin{abstract}
We have analyzed the two-dimensional antiferromagnetic Heisenberg model on the
triangular lattice using a Schwinger boson mean-field theory.
By expanding around a state with local $120^\circ$ order,
we obtain, in the limit of infinite spin, results for the excitation spectrum
in complete agreement with linear spin wave theory (LSWT).
In contrast to LSWT, however, the modes at the ordering wave vectors acquire
a mass for finite spin. We discuss the origin of this effect.
\end{abstract}

\vspace{4mm}


\narrowtext
\section*{Introduction}
The two-dimensional spin-$\frac{1}{2}$ Heisenberg antiferromagnet (HAFM) has
received much
attention during the last few years. The main reason for this new interest
stems from the discovery of the ceramic superconductors, which has led to
extensive studies of the HAFM on a square lattice~\cite{Manousakis1991}.
An important focal point has been the experimentally observed competition
between superconductivity and antiferromagnetic order in these
materials. This has encouraged
numerous investigations of disordering effects, in particular
the role of frustrating interactions.

To elucidate the role of
frustration, it is interesting to study cases where frustration is
intrinsic to the lattice structure such as
the HAFM on a triangular or Kagom\'e lattice.
Early papers on the triangular HAFM~\cite{PWAnd1973,KalLau1987}
suggested that the spin-$\frac{1}{2}$ ground state is a spin liquid,
lacking long-range order (LRO).
More recent studies~\cite{HuseElser1988,JoliLeGuill1989,%
YoshMiya1991,SingHuse1992,BerLhuPie1992,ElsSingYou1993}
point to the scenario of an ordered ground state possibly quite close to a
disordering
transition. Yet other studies support the original proposal of a
disordered ground state~\cite{Wang1992,YanWarGir1993}.

The spin dynamics of the model,
intimately linked to the
character of the ground state, also remains an open question.
The problem of the spin dynamics above a ground state with N\'eel-type
order (a $120^\circ$ state) has been attacked with a variety of methods, like
linear spin-wave theory (LSWT)~\cite{JoliLeGuill1989},
nonlinear $\sigma$ model~\cite{DomRead1989} and
Schwinger-boson mean-field theory (SBMFT)~\cite{YoshMiya1991,LefHed1994}.
The SBMFT has proved
successful in incorporating quantum fluctuations \cite{AA}, but
the choice of mean-field parameters
is a delicate
matter~\cite{CommentonChanColeLark1991,ReplytoCommentonChanColeLark1991}.
Yoshioka and Miyazaki~\cite{YoshMiya1991} have used the SBMFT combined
with a resonant valence bond ansatz,
and a similar theory including a ferromagnetic component in the mean-field
expansion is recently analyzed by Lefmann and
Hedeg\aa rd~\cite{LefHed1994}.
A drawback of these applications of SBMFT is that the LSWT structure
of the dispersion relation cannot be recovered in the limit of infinite spin.
Specifically, only two massless spin wave modes can be obtained, in contrast to
the three massless modes of LSWT.
This is a weakness, since, at an analytical level, the only reliable way
to motivate the validity of
the theory is to use an approach that correctly reproduces the limit
of classical magnetism~\cite{ReplytoCommentonChanColeLark1991}.

We here employ another mean-field treatment, where we use a
spin-rotation to identify the relevant mean-field parameters,
treating the spin dynamics above a state close to the classical ground state.
In this way {\em three} modes are obtained, two of which exhibit a massgap at
finite spin. However, in the limit of infinite spin, the gap asymptotically
vanishes on the relevant energy scale and the corresponding spin wave
velocities exactly approach those of LSWT. We discuss the origin of
massgaps within this version of SBMFT and "explain"
why there are no gaps in LSWT at any value of the spin.

\section*{The model}

We consider the Hamiltonian
\begin{equation}
{\cal H} =
J_{1}\sum_{\bbox{R},\bbox{\alpha}}{\bf S}_{\bbox{R}} \cdot
  {\bf S}_{\bbox{R}+\bbox{\alpha}} \ \  ,
\end{equation}
defined on a triangular lattice and with $J_{1} \geq 0$.
The sums run over all lattice sites $\bbox{R}$ and three of the six
nearest-neighbor vectors $\bbox{\alpha}$. We choose the vectors
$\bbox{\alpha}$ to be
\begin{mathletters}
\label{alpha}
\begin{eqnarray}
\bbox{\alpha}_1 & = & - \frac{1}{2} {\bf e}_x - \frac{\sqrt{3}}{2} {\bf e}_y
 \ \ , \\
\bbox{\alpha}_2 & = & - \frac{1}{2} {\bf e}_x + \frac{\sqrt{3}}{2} {\bf e}_y
 \ \ , \\
\bbox{\alpha}_3 & = & {\bf e}_x\ \ ,
\end{eqnarray}
\end{mathletters}
in units where the lattice constant equals 1 (see Fig.~\ref{fig:triangular}).

The triangular lattice is a Bravais lattice and
can be divided into three equal interlacing sublattices (tripartite lattice).
In Fig.~\ref{fig:triangular} the three sublattices are indicated.
It follows that in the classical limit of large spin, the
ground state is a $120^\circ$ state, i.e.
a coplanar state with spins on each sublattice being
ferromagnetically ordered and rotated $120^\circ$ vis a vis the spins
on the two other sublattices. (The lattice constant of a sublattice is
$\sqrt3$ times the nearest neighbor spacing why this state also is
called a "$\sqrt3 \times \sqrt3$ state".)

When expanding around such a $120^\circ$ state, it is convenient
to start with a
local rotation of the spin coordinate system.
The spins on neighboring sublattices are rotated by $+120^\circ$ around the
$y$-axis when going from $\bbox{R}$ to $\bbox{R}+\bbox{\alpha}$
\footnote{To go in the $+\bbox{\alpha}$ and in
the $-\bbox{\alpha}$ directions is not equivalent since the $120^\circ$
state breaks parity. Note that in
two dimensions a parity transformation is defined by a reflection in a line.}:
\begin{mathletters}
\label{eqn:Spinrot}
\begin{eqnarray}
S^x_{\bbox{R}} & \rightarrow &
\cos(\bbox{Q} \cdot \bbox{R}) S^x_{\bbox{R}} +
\sin(\bbox{Q} \cdot \bbox{R}) S^z_{\bbox{R}} \ \ ,\\
S^y_{\bbox{R}} & \rightarrow & S^y_{\bbox{R}} \ \ ,\\
S^z_{\bbox{R}} & \rightarrow &
\cos(\bbox{Q} \cdot \bbox{R}) S^z_{\bbox{R}} -
\sin(\bbox{Q} \cdot \bbox{R}) S^x_{\bbox{R}} \ \ ,
\end{eqnarray}
\end{mathletters}
where $\bbox{Q}$ is defined by
$\bbox{Q} \cdot \bbox{\alpha} = \frac{2\pi}{3}+2\pi n$ with $n$ integer.
This implies that
\begin{eqnarray}
\label{eqn:SpinrotHam}
{\bf S}_{\bbox{R}} \cdot {\bf S}_{\bbox{R}+\bbox{\alpha}}
& \rightarrow & \frac{\sqrt{3}}{2} \left( S_{\bbox{R}}^{z}
S_{\bbox{R}+\bbox{\alpha}}^{x} - S_{\bbox{R}}^{x}
S_{\bbox{R}+\bbox{\alpha}}^{z} \right)
- \frac{1}{2} \left( S_{\bbox{R}}^{z}
S_{\bbox{R}+\bbox{\alpha}}^{z} + S_{\bbox{R}}^{x}
S_{\bbox{R}+\bbox{\alpha}}^{x} \right)
+ S_{\bbox{R}}^{y} S_{\bbox{R}+\bbox{\alpha}}^{y}
\end{eqnarray}

Let us pick a specific $120^\circ$ state, with the (unrotated) spins on the
first sublattice
pointing in the $z$-direction while on the other two sublattices
they span
the $x-z$ plane with directions $-120^\circ$ ($+120^\circ$) with respect to
the $z$-axis when
going in the $\bbox{\alpha}$ ($-\bbox{\alpha}$) direction in the lattice.
Any classical
ground state can clearly be reached from this state by a (global) $SO(3)$
rotation of the spins.
We immediately see that the spin rotation in Eqs.~(\ref{eqn:Spinrot})
transfers this $120^\circ$  state into a ferromagnetic state with
quantization axes lying in the $x-z$ plane.
The first term in Eq.~(\ref{eqn:SpinrotHam}) vanishes when acting
on a ferromagnetic state, and we may hence neglect it when modeling
states with a local structure close to a $120^\circ$ state.
In the rotated basis, the effective Hamiltonian is thus given by
\begin{equation}
{\cal H} =
J_{1}\sum_{\bbox{R},\bbox{\alpha}}-\frac{1}{2}\left(S^x_{\bbox{R}}
S^x_{\bbox{R}+\bbox{\alpha}} +
S^z_{\bbox{R}}S^z_{\bbox{R}+\bbox{\alpha}}\right)
+S^y_{\bbox{R}}S^y_{\bbox{R}+\bbox{\alpha}}\ \ .
\label{eqn:effHam}
\end{equation}
We have verified that in our case this approach is equivalent
to that in
Ref.~\onlinecite{ChanColeLark1990} where an analysis of the frustrated model
on the square lattice is performed.
It is interesting to note that in LSWT the term that corresponds
to $\frac{\sqrt{3}}{2} \left( S_{\bbox{R}}^{z}%
S_{\bbox{R}+\bbox{\alpha}}^{x} - S_{\bbox{R}}^{x}%
S_{\bbox{R}+\bbox{\alpha}}^{z} \right)$
in Eq.~(\ref{eqn:SpinrotHam}) is also
neglected since it is cubic in the Holstein-Primakoff bosons.

The Schwinger-boson calculation that follows is similar to
the treatment of the $J_1-J_2$ model on the square lattice ~\cite{ToPeHe}
and on the honeycomb lattice~\cite{AnPeTo}.
The spin operators ${\bf S}_{\bbox{R}}$ at each lattice site are replaced by
two species of Schwinger bosons $ b^{\dagger}_{\mu \bbox{R}} \ (\mu = 1,2)$ via
\begin{equation}
{\bf S}_{\bbox{R}} = {1 \over 2}  b^{\dagger}_{\mu \bbox{R}}
{\mbox{\boldmath $\sigma$}}^{\mathop{\phantom{\dagger}}}_{\mu \nu}
  b^{\mathop{\phantom{\dagger}}}_{\nu \bbox{R}} \ \ \ , \label{julianS}
\end{equation}
with the local constraints
$ b^{\dagger}_{\mu {\bbox{R}} }
  b^{\mathop{\phantom{\dagger}}}_{\mu {\bbox{R}} } = 2S.$
Here $\mbox{\boldmath $\sigma$} = ({\sigma}^x, {\sigma}^y, {\sigma}^z)$
is the vector of Pauli matrices,
and summation over repeated (Greek) indices is implied.
This leads us to the following Hamiltonian:
\begin{equation}
{\cal H} =
  - J_1 \sum_{\bbox{R}, \bbox{\alpha}} \left( - \frac{3}{4}
{\cal W}^{\cal A}_{\bbox{R},\bbox{\alpha}}
  + \frac{1}{4} {\cal W}^{\cal B}_{\bbox{R},\bbox{\alpha}} \right)
 + \sum_{\bbox{R} } \lambda_{\bbox{R}} [ b^{\dagger}_{\mu  \bbox{R}}
 b^{\mathop{\phantom{\dagger}}}_{\mu \bbox{R}} -2S] \ \ ,
\label{bondH}
\end{equation}
where we have included the local constraints with Lagrange
multipliers $\lambda_{\bbox{R}}$ at each site.
The expression for the summands is
\begin{equation}
{\cal W}^{\cal X}_{\bbox{R},\bbox{\alpha}} \ \  =
   \ \  \mbox{$\frac{1}{2}$} : {\cal X}^{\dagger}_{\bbox{R}, \bbox{\alpha}}
 {\cal X}^{\mathop{\phantom{\dagger}}}_{\bbox{R}, \bbox{\alpha}}: - S^2 \ \ ,
\label{Wdef}
\end{equation}
with ${\cal X}^{\mathop{\phantom{\dagger}}}_{\bbox{R}, \bbox{\alpha}}$ any of
 the two link operators
\begin{mathletters}
\begin{eqnarray}
{\cal A}^{\mathop{\phantom{\dagger}}}_{\bbox{R}, \bbox{\alpha}} & \equiv &
 b^{\mathop{\phantom{\dagger}}}_{1 \bbox{R}}
 b^{\mathop{\phantom{\dagger}}}_{1 \bbox{R}+ \bbox{\alpha}} +
 b^{\mathop{\phantom{\dagger}}}_{2 \bbox{R}}
 b^{\mathop{\phantom{\dagger}}}_{2 \bbox{R} + \bbox{\alpha}} \ ,
 \label{bondOPa} \\
  {\cal B}^{\mathop{\phantom{\dagger}}}_{\bbox{R}, \bbox{\alpha}} & \equiv &
  b^{\dagger}_{1 \bbox{R}}  b^{\mathop{\phantom{\dagger}}}_{1 \bbox{R} +
 \bbox{\alpha}} +
 b^{\dagger}_{2 \bbox{R}}  b^{\mathop{\phantom{\dagger}}}_{2 \bbox{R} +
 \bbox{\alpha}} \ . \label{bondOPb}
\end{eqnarray}
\end{mathletters}

The mean-field theory is finally generated by the Hartree-Fock decoupling
\begin{equation}
:\! {\cal X}^{\dagger}_{\bbox{R}, \bbox{\alpha}}
 {\cal X}^{\mathop{\phantom{\dagger}}}_{\bbox{R}, \bbox{\alpha}}\!:
 \,\rightarrow
 {\cal X}^{\dagger}_{\bbox{R}, \bbox{\alpha}} \langle
 {\cal X}^{\mathop{\phantom{\dagger}}}_{\bbox{R}, \bbox{\alpha}} \rangle +
\langle  {\cal X}^{\dagger}_{\bbox{R}, \bbox{\alpha}}
 \rangle {\cal X}^{\mathop{\phantom{\dagger}}}_{\bbox{R}, \bbox{\alpha}} -
\langle  {\cal X}^{\dagger}_{\bbox{R}, \bbox{\alpha}} \rangle \langle
 {\cal X}^{\mathop{\phantom{\dagger}}}_{\bbox{R}, \bbox{\alpha}} \rangle
\ \ ,\label{Hartree}
\end{equation}
where the link fields (mean-field parameters)
$Q_1 \equiv \langle{\cal A}^{\mathop{\phantom{\dagger}}}_{\bbox{R} ,
 \bbox{\alpha}}\rangle$ and
$Q_2 \equiv \langle {\cal B}^{\mathop{\phantom{\dagger}}}_{\bbox{R} ,
 \bbox{\alpha}}\rangle$
are taken to be uniform and real.
In our mean-field treatment, we also replace the local
Lagrange multipliers ${\lambda}_{\bbox{R}}$ by a single parameter $\lambda $.

For comparison of our results with those of
LSWT~\cite{JoliLeGuill1989} it is convenient to Fourier-transform the
 Schwinger
bosons independently on each
sublattice,
\begin{equation}
 b^{\mathop{\phantom{\dagger}}}_{\mu {\bbox{R}_m}} = \frac{1}{\sqrt{N/2}}
 \sum_{\bbox{k}}
e^{-i \bbox{k} \cdot \bbox{R}_m}
a^{\mathop{\phantom{\dagger}}}_{m \mu \bbox{k}} \ \  ,
\ \  \mbox{$m=1,2,3$}\ \ ,
\label{Ftransf}
\end{equation}
with $\bbox{R}_1$, $\bbox{R}_2$ and $\bbox{R}_3$ on the three different
sublattices respectively.
We then need to decouple the different bosons on the sublattices.
This is done
by the canonical transformations
\begin{mathletters}
\begin{eqnarray}
a^{\mathop{\phantom{\dagger}}}_{1 \mu \bbox{k}} &=&
\frac{1}{\sqrt{3}}\left(\beta^{\mathop{\phantom{\dagger}}}_{1 \mu \bbox{k}} +
\beta^{\mathop{\phantom{\dagger}}}_{2 \mu \bbox{k}} +
\beta^{\mathop{\phantom{\dagger}}}_{3 \mu \bbox{k}} \right) \, , \\
a^{\mathop{\phantom{\dagger}}}_{2 \mu \bbox{k}} &=&
\frac{1}{\sqrt{3}}\left(\beta^{\mathop{\phantom{\dagger}}}_{1 \mu \bbox{k}} +
j^* \,\beta^{\mathop{\phantom{\dagger}}}_{2 \mu \bbox{k}} +
j \,\beta^{\mathop{\phantom{\dagger}}}_{3 \mu \bbox{k}} \right) \, , \\
a^{\mathop{\phantom{\dagger}}}_{3 \mu \bbox{k}} &=&
\frac{1}{\sqrt{3}}\left(\beta^{\mathop{\phantom{\dagger}}}_{1 \mu \bbox{k}} +
j \,\beta^{\mathop{\phantom{\dagger}}}_{2 \mu \bbox{k}} +
j^* \,\beta^{\mathop{\phantom{\dagger}}}_{3 \mu \bbox{k}} \right) \, ,
\end{eqnarray}
\end{mathletters}
where $j=\exp(i\frac{2 \pi}{3})$.

Finally we use the Bogoliubov transformations
\begin{mathletters}
\label{eqn:Boug}
\begin{eqnarray}
\beta^{\mathop{\phantom{\dagger}}}_{1 \mu \bbox{k}} & = &
 \cosh{{\vartheta}_{1 \bbox{k}}}
 c^{\mathop{\phantom{\dagger}}}_{1 \mu \bbox{k}}
+ \sinh{{\vartheta}_{1 \bbox{k}}}  c^{\dagger}_{1 \mu -\bbox{k}} \ \  ,
\label{Btransfa} \\
\beta^{\mathop{\phantom{\dagger}}}_{2 \mu \bbox{k}} & = &
 \cosh{{\vartheta}_{2 \bbox{k}}}
 c^{\mathop{\phantom{\dagger}}}_{2 \mu \bbox{k}}
+ \sinh{{\vartheta}_{2 \bbox{k}}}  c^{\dagger}_{3 \mu -\bbox{k}} \ \  ,
\label{Btransfb} \\
\beta^{\mathop{\phantom{\dagger}}}_{3 \mu \bbox{k}} & = &
 \cosh{{\vartheta}_{3 \bbox{k}}}
c^{\mathop{\phantom{\dagger}}}_{3 \mu \bbox{k}}
+ \sinh{{\vartheta}_{3 \bbox{k}}}  c^{\dagger}_{2 \mu -\bbox{k}} \ \  ,
\label{Btransfc}
\end{eqnarray}
\end{mathletters}
with
\begin{equation}
{\rm tanh} (2{\vartheta}_{m \bbox{k}}) =
\frac{\mbox{$\frac{9}{2}$} J_1 Q_1  \gamma_{m,\bbox{k}}}
{\lambda + \mbox{$\frac{3}{2}$} J_1 Q_2 \gamma_{m,\bbox{k}} } \ \ ,
\ \  \mbox{$m=1,2,3$} \ \ ,
\label{angle}
\end{equation}
to diagonalize the resulting Hamiltonian.
This yields free bosons
$c_{m \mu \bbox{k}}$ with dispersion relations
\begin{equation}
\omega_{m \bbox{k}} = \sqrt{(\lambda + \mbox{$\frac{3}{2}$} J_1 Q_2
\gamma_{m,\bbox{k}} )^2 -
(\mbox{$\frac{9}{2}$} J_1 Q_1 \gamma_{m,\bbox{k}} )^2}
\ \  , \ \  \mbox{$m=1,2,3$} \ \ ,
\label{dispersion1}
\end{equation}
where the geometrical factors are given by
\begin{mathletters}
\label{eqn:geo.fact}
\begin{eqnarray}
\gamma_{1, \bbox{k}} & = & \frac{1}{6}
\sum_{\bbox{\alpha}} \cos ( \bbox{k} \cdot \bbox{\alpha} ) \ \ ,\\
\gamma_{2, \bbox{k}} & = & \frac{1}{6}
\sum_{\bbox{\alpha}} \cos ( \bbox{k} \cdot \bbox{\alpha} + \frac{2 \pi}{3} )
 \ \ ,\\
\gamma_{3, \bbox{k}} & = & \frac{1}{6}
\sum_{\bbox{\alpha}} \cos ( \bbox{k} \cdot \bbox{\alpha} - \frac{2 \pi}{3} )
 \ \ .
\end{eqnarray}
\end{mathletters}

By minimizing the free energy we obtain the three equations needed to
determine the mean-field parameters:
\begin{mathletters}
\label{INTEQ} 
\begin{eqnarray}
\frac{1}{3} \sum_{m=1}^3 \int \frac{d^2 k}{A} \cosh
 (2 {\vartheta}_{m \bbox{k}})
  (n_{m \bbox{k} }+ \mbox{$\frac{1}{2}$}) - (S+\mbox{$\frac{1}{2}$})  &=& 0
 \ \ ,\label{INTEQa}\\
\frac{1}{3} \sum_{m=1}^3 \int \frac{d^2 k}{A} \sinh(2 {\vartheta}_{m \bbox{k}}
 )
\gamma_{m, \bbox{k}}(n_{m \bbox{k} } + \mbox{$\frac{1}{2}$}) -
 \mbox{$\frac{1}{2}$}
 Q_1  &= &  0 \ \  ,   \label{INTEQb}\\
\frac{1}{3} \sum_{m=1}^3 \int \frac{d^2 k}{A}
\cosh(2 {\vartheta}_{m \bbox{k}} )
\gamma_{m, \bbox{k}}(n_{m \bbox{k} } + \mbox{$\frac{1}{2}$}) -
 \mbox{$\frac{1}{2}$}
 Q_2  &= &  0 \ \  .   \label{INTEQc}
\end{eqnarray}
\end{mathletters}
Here $A=8 \pi^{2}/(3 \sqrt{3})$ is the area of the reciprocal unit cell of
a sublattice, and $n_{m \bbox{k} }= [\exp(\beta\omega_{m \bbox{k}})-1]^{-1}$
is the Bose occupation number.
Solving the mean-field equations numerically yields values for
$Q_1$, $Q_2$, and $\lambda$, which are used to determine
thermodynamic quantities at finite temperatures.

Given a Bravais lattice, as in the present case, a division
into sublattices is not necessary when Fourier-transforming
(as it is on a non-Bravais lattice) and the theory may be formulated
in terms of only one set of bosons.
We have chosen to transform each sublattice independently in order
to make the comparison
with LSWT results more transparent.
In our treatment we thus have a small Brillouin zone
and three sets of bosons ({\em picture A}), which is equivalent to
{\em picture B} where the larger Brillouin zone
of the real lattice and one of the sets of bosons
(say $c^{\mathop{\phantom{\dagger}}}_{1 \mu \bbox{k}}$) is used
(see Fig.~\ref{fig:Brill}).
The equivalence
of the two pictures  makes it evident why
$c^{\mathop{\phantom{\dagger}}}_{2 \mu \bbox{k}}$ has to be combined with
$c^{\mathop{\phantom{\dagger}}}_{3 \mu -\bbox{k}}$ in Eqs.~(\ref{Btransfb})
and~(\ref{Btransfc}).
It also explains the
structure of the geometrical factors in Eq.~(\ref{eqn:geo.fact})
(remembering that $\bbox{Q}\cdot\bbox{\alpha}=\frac{2\pi}{3}$).
Until now we have used picture A, but e.\ g.\ in the figures
we will use picture B (one mode) instead of showing plots for
three different modes. Since mode $2$ and $3$ in picture A are
degenerate we will sometimes refer to them as the corner modes which
is picture B language. To avoid confusion we indicate
in the following which picture we use.

\section*{Zero temperature formalism}
In two dimensions there can be LRO only at zero temperature. This
means that exactly at $T=0$ there is an abrupt phase transition,
and we must be careful when reducing $T$ to zero.
As we see from Eq.~(\ref{INTEQ}), the parameters $Q_1$, $Q_2$ and $\lambda$
will depend on temperature via the Bose occupation number $n_{m \bbox{k}}$,
and hence, to calculate properties at $T=0$ we need to
pass to a zero temperature
formalism. Following
Ref.~\onlinecite{Taka1989} we note that LRO
corresponds to a condensation of the Schwinger bosons at some points
$\bbox{K}_m$ in the Brillouin zone and we obtain the
mean-field equations at $T=0$ from Eq.~(\ref{INTEQ}) by the
replacement
\begin{equation} \label{eqn:T=0-replacement}
\frac{n_{m \bbox{k}}}{\omega_{m \bbox{k}}} \rightarrow S_m^{*}\delta^{(2)}
(\bbox{k}-\bbox{K}_m) \ \ ,
\end{equation}
where $S_m^{*}$ is a new unknown quantity measuring the Bose condensate.
Since the condensation of the bosons implies that
there is no gap in the spin-wave spectrum ($\omega_{m \bbox{k}=\bbox{K}_m}=0$
in (\ref{dispersion1})) at $T=0$, these $m$ equations together with the
three equations in~(\ref{INTEQ}) are sufficient to determine the
$3+m$ parameters $\lambda$, $Q_1$, $Q_2$ and $S_m^{*}$.
The procedure is the following:
We start at a finite temperature and use Eqs.~(\ref{INTEQ}) to
determine the mean-field parameters
which gives us the dispersion relation. We then lower the temperature and
examine the dispersion relation to identify the
points in the Brillouin zone where the gap scales down with temperature.
In our case this
happens for each of the three modes (picture A) at
$\bbox{k} = \bbox{0}$.
We expand the dispersion relation around these points and
for small $\bbox{k}$ the dispersion relations (\ref{dispersion1}) takes the
relativistic form
 ${\omega}_{m \bbox{k} } = c_m \sqrt{(M_mc_m)^2 + |{\bbox{k}}|^2}$.
Carrying out the expansion we find that the spin-wave
velocities $c_m$ are given by
\begin{equation}
c_1 = \sqrt{(\mbox{$\frac{9}{4}$} J_1 Q_1)^2 \mbox{$\frac{1}{2}$} -
\mbox{$\frac{3}{8}$}
J_1 Q_2 (\lambda + \mbox{$\frac{3}{4}$} J_1 Q_2 )}
\ \  , \label{c1}
\end{equation}
and
\begin{equation}
c_2=c_3=\sqrt{(\mbox{$\frac{9}{8}$} J_1 Q_1)^2 \mbox{$\frac{1}{2}$} +
\mbox{$\frac{3}{16}$}
J_1 Q_2 (\lambda - \mbox{$\frac{3}{8}$} J_1 Q_2 )}
\ \  . \label{c23}
\end{equation}
The masses $M_m$ in the energy gap $\Delta_m=M_mc_m^2$ of the
spin-wave excitations can in turn be extracted from
\begin{equation}
\Delta_1 = \sqrt{(\lambda + \mbox{$\frac{3}{4}$}J_1 Q_2 )^2 -
(\mbox{$\frac{9}{4}$} J_1 Q_1)^2} \ \ ,
\label{M1}
\end{equation}
and
\begin{equation}
\Delta_2=\Delta_3= \sqrt{(\lambda - \mbox{$\frac{3}{8}$}J_1 Q_2 )^2 -
(\mbox{$\frac{9}{8}$} J_1 Q_1)^2} \ \ .
\label{M23}
\end{equation}
To examine whether we have condensation of bosons for all modes, we first
consider the gaps, or the masses, since condensation of the bosons at
$\bbox{K}_m=\bbox{0}$ implies that we have no gap in the spectrum
at this point.
To get all modes  massless ($\Delta_1 = \Delta_2=\Delta_3= 0$) we need
to have $Q_1=Q_2=\frac{2}{3}\frac{\lambda}{J_1}$ (from Eqs.~(\ref{M1})
and~(\ref{M23})). Since $\gamma_{1,\bbox{0}}=\frac{1}{2}$ and
 $\gamma_{2,\bbox{0}}=\gamma_{3,\bbox{0}}=
-\frac{1}{4}$ we see from Eq.~(\ref{angle}) that we must have equalities
between {\em finite} quantities as
$\omega_{1 \bbox{k}}\sinh(2\vartheta_{1 \bbox{0}})=
\omega_{1 \bbox{k}}\cosh(2\vartheta_{1 \bbox{0}})$ and
$\omega_{2 \bbox{k}}\sinh(2\vartheta_{2  \bbox{0}})=
\omega_{3 \bbox{k}}\sinh(2\vartheta_{3 \bbox{0}})
=-\omega_{2 \bbox{k}}\cosh(2\vartheta_{2 \bbox{0}})
=-\omega_{3 \bbox{k}}\cosh(2\vartheta_{3 \bbox{0}})$.
But doing the replacement in Eq.~(\ref{eqn:T=0-replacement}) into
Eq.~(\ref{INTEQ})
for all three modes then
gives $Q_1 \neq Q_2$ from
Eqs.~(\ref{INTEQb}) and~(\ref{INTEQc}),
in contradiction to the initial assumption. The only way to get three
massless modes is
to have $\omega_{1 \bbox{k}}\sinh(2\vartheta_{1 \bbox{0}})=
\omega_{1 \bbox{k}}\cosh(2\vartheta_{1 \bbox{0}})$ {\em infinite}, which then
yields $Q_1 =Q_2$ at {\em infinite spin $S$} as seen from Eq.~(\ref{INTEQa}).

The result of the procedure above is that for finite
spin $S$ we will get a condensation of the bosons only
for the first mode which scales faster to zero than the two other modes.
For this mode we do the replacement shown in
Eq.~(\ref{eqn:T=0-replacement}) ($m=1$ and $\bbox{K}_1=\bbox{0}$)
into Eq.~(\ref{INTEQ}) ($S_2^{*}=S_3^{*}=0$) and
since $\Delta_1=0$, Eq.~(\ref{M1}) gives $\lambda$ in terms of $Q_1$ and $Q_2$.
After these substitutions we have no temperature dependence in
Eqs.~(\ref{INTEQ}).

With this zero temperature formalism we have studied the
{\em spin} dependence of the mass $M_2$ ($=M_3$).
As we see in Fig.~\ref{fig:M2vsS} the mass goes to zero in
the limit of infinitely large spin and we recover
the three massless modes of LSWT.
The dispersion relation in Eq.~(\ref{dispersion1}) can be rewritten as
\begin{equation}
\frac{\omega_{m \bbox{k}}}{J_1} = \sqrt{(\mbox{$\frac{\lambda}{J_1}$} +
\mbox{$\frac{3}{2}$} ( 3 Q_1 + Q_2 )
\gamma_{m,\bbox{k}})(\mbox{$\frac{\lambda}{J_1}$} - \mbox{$\frac{3}{2}$}
( 3 Q_1 -Q_2 ) \gamma_{m,\bbox{k}})} \ \ ,
\label{dispersion2}
\end{equation}
which immediately can be compared with
the LSWT result,
Eq.~(13) in Ref.~\onlinecite{JoliLeGuill1989}.
To get the LSWT results we need $Q_1=Q_2=\frac{2}{3}\frac{\lambda}{J_1}$,
which is (as we have already seen) the condition to obtain three massless modes
($\Delta_1 = \Delta_2=\Delta_3= 0$).
Thus, in the limit of infinitely large spin, where we in fact
obtain three massless modes,
we {\em exactly} recover the LSWT results.
In Fig.~\ref{fig:Omega} we show the dispersion relation for two
values of the spin $S$. We see that on the relevant energy scale
the gaps at the corners (picture B) go to zero and for infinite
spin we recover the LSWT dispersion relation.

\begin{figure}
\caption{The ratio between the center mode spin-wave velocity $c_1$ and
the corner mode (picture B (see text)) spin-wave velocities
$c_2$ ($=c_3$) vs. spin $S$
compared with the LSWT result $\protect{\sqrt{2}}$. \label{fig:c1c23ratiovsS}}
\end{figure}
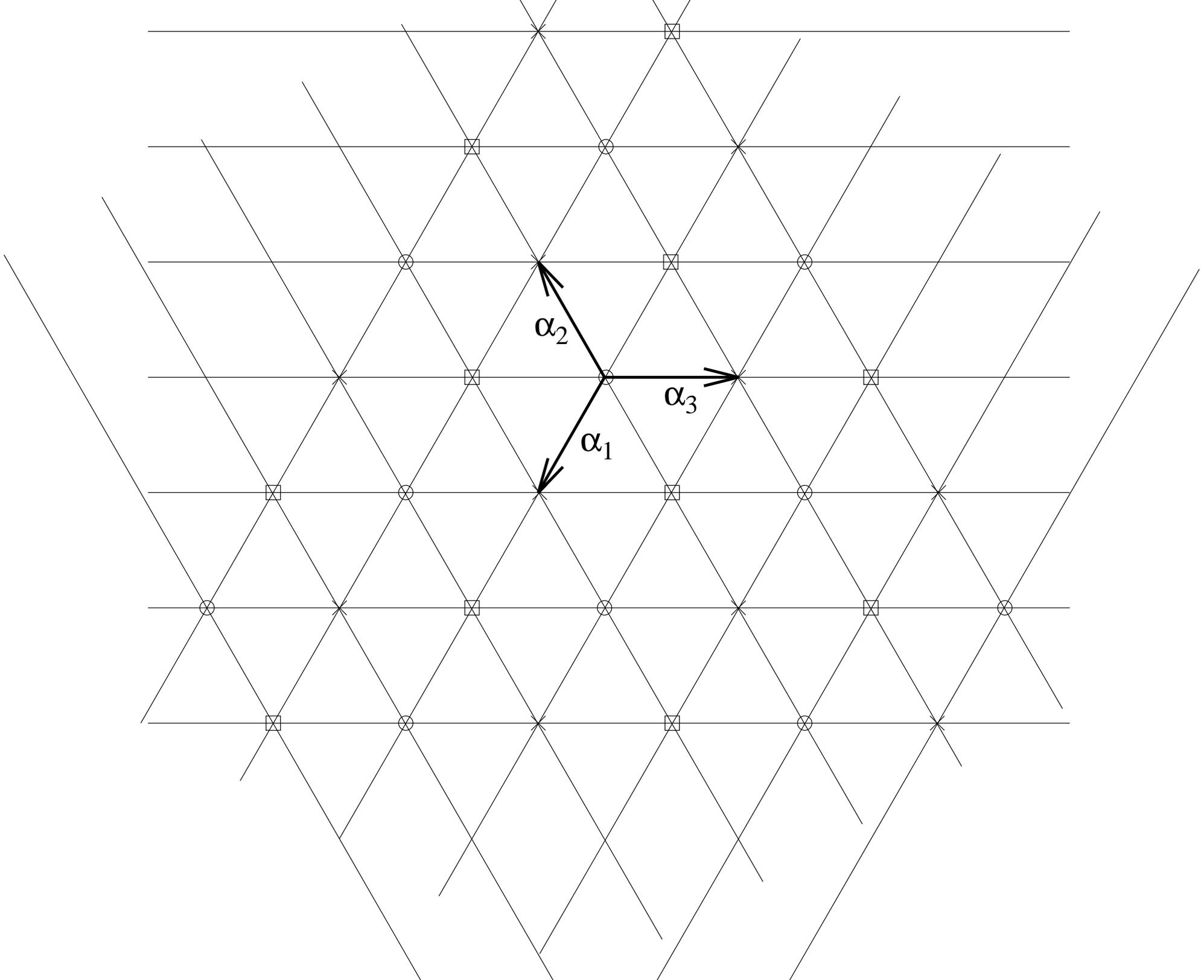
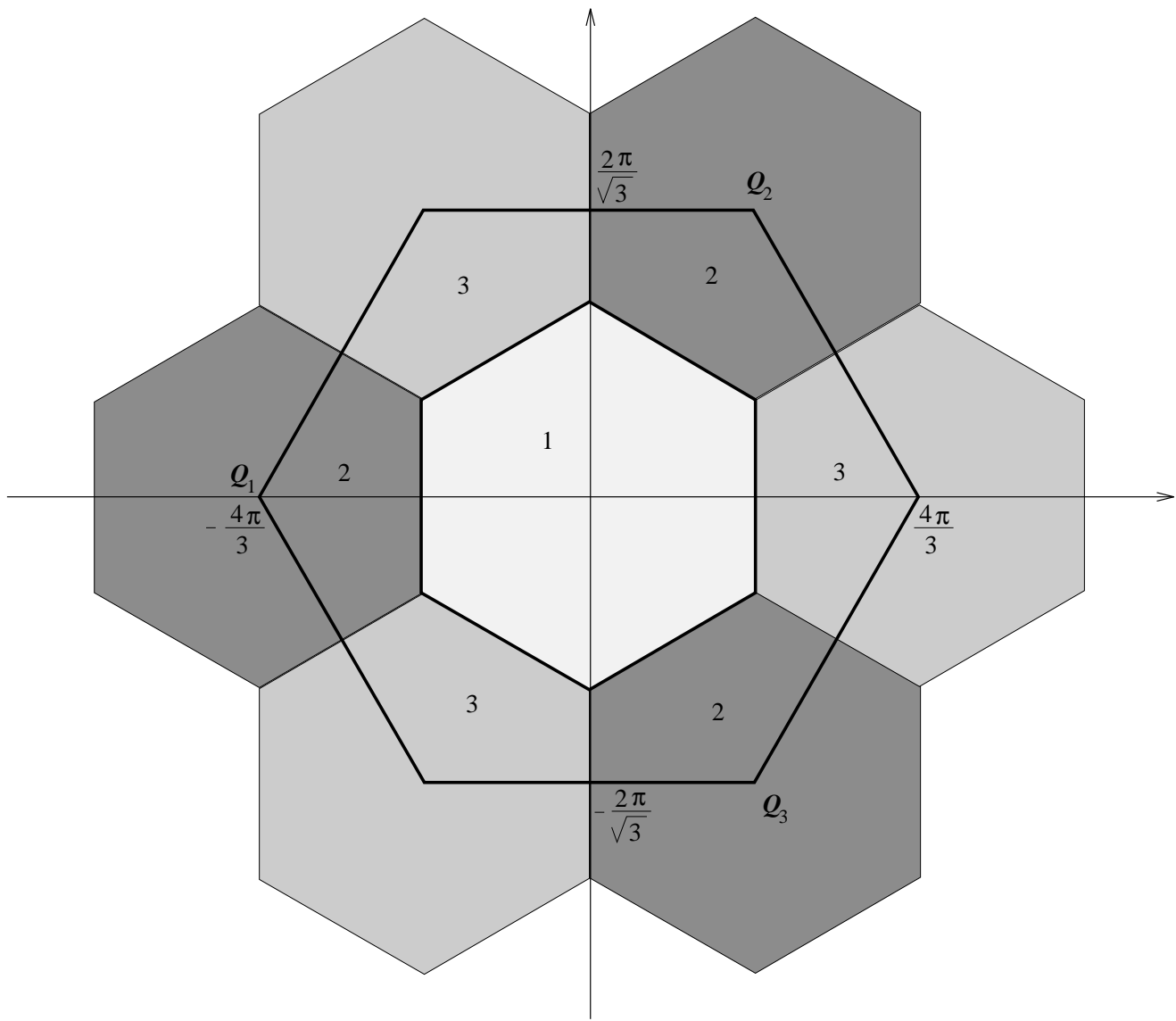
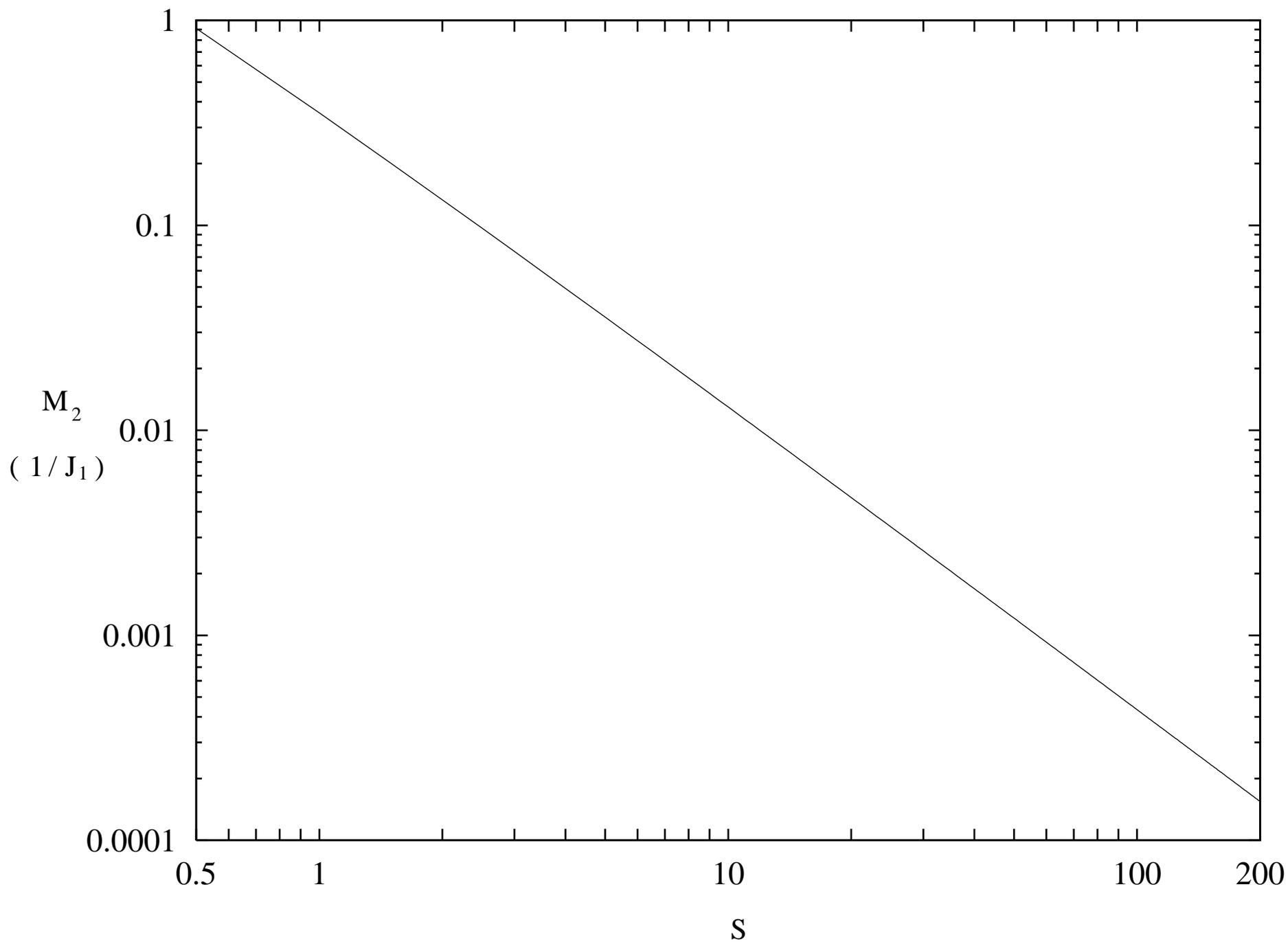
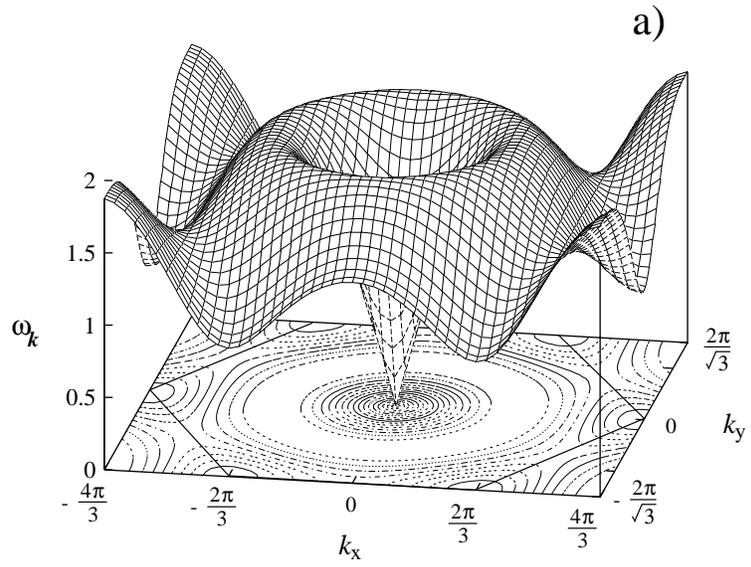
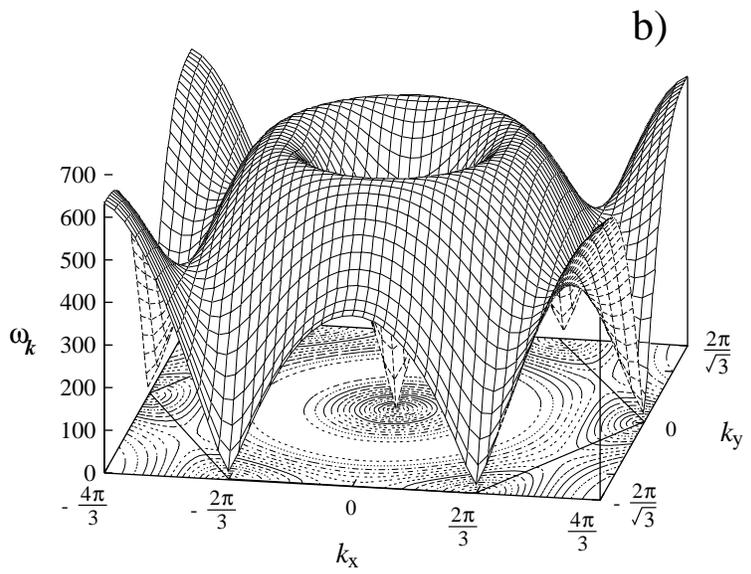
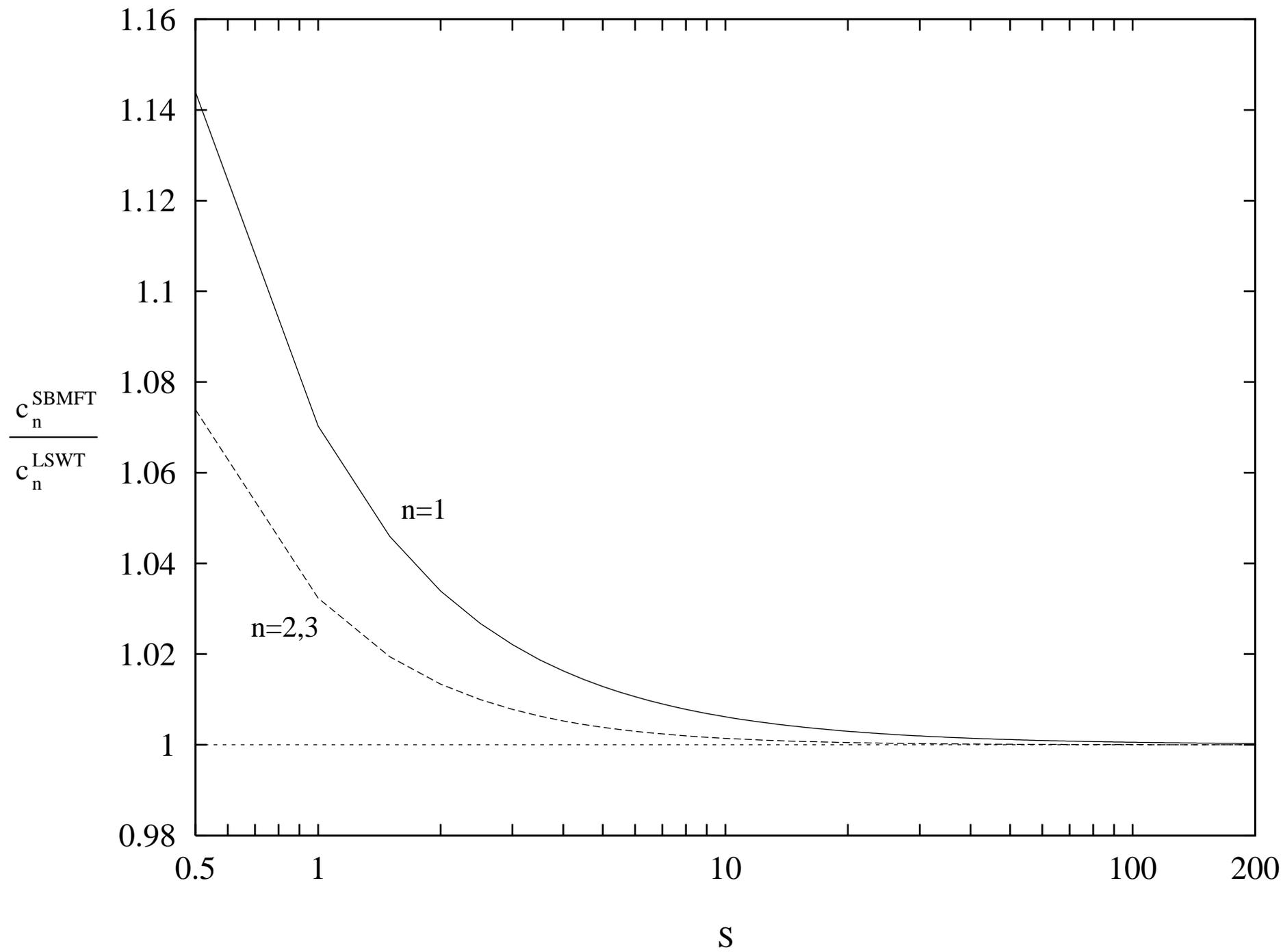
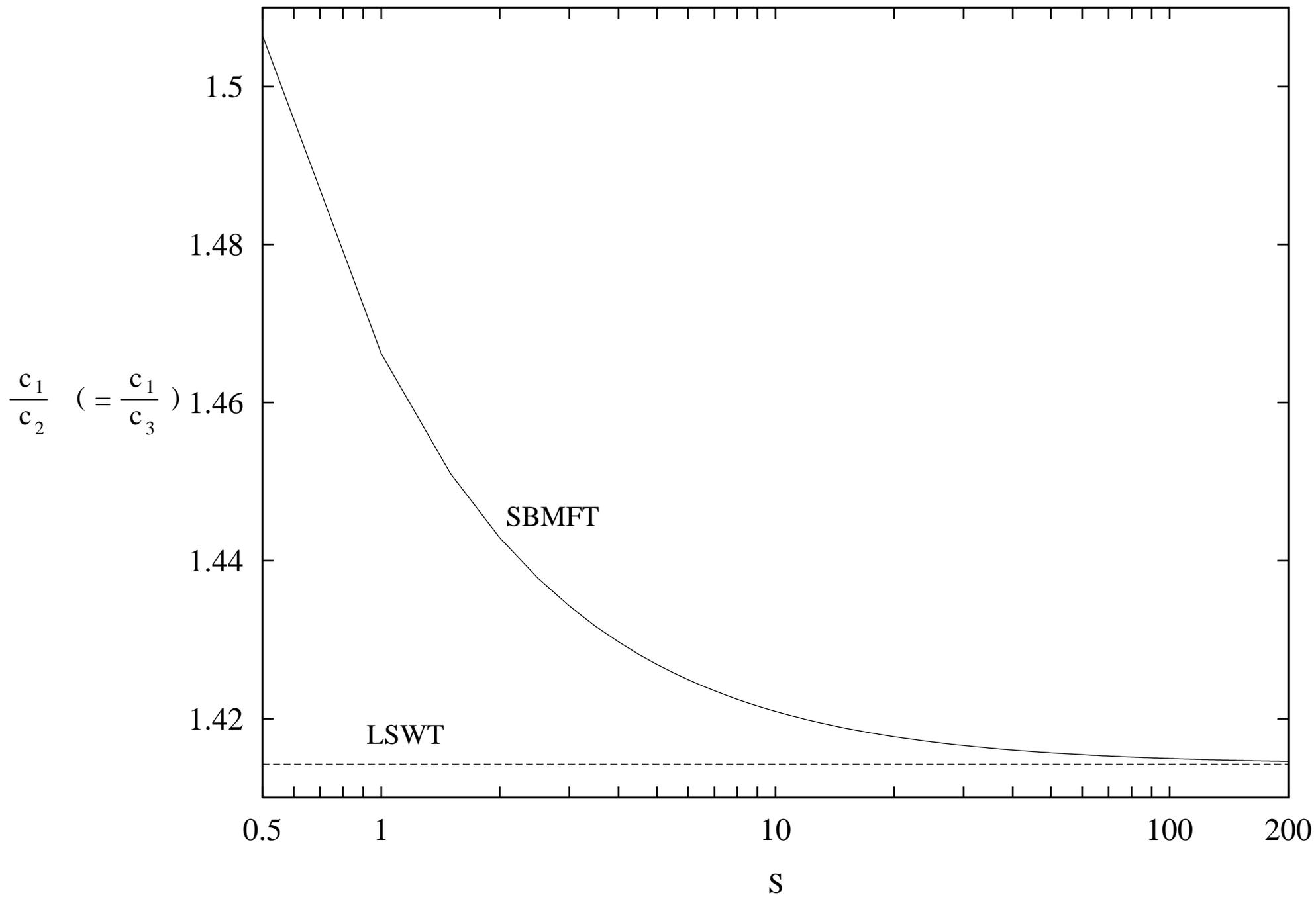

We have also calculated the spin-wave velocities $c_m$ ($m=1,2,3$)
at $T=0$. In Fig.~\ref{fig:ratioc1ratioc23vsS} we see
the results compared with the LSWT results extracted from
Ref.~\onlinecite{JoliLeGuill1989} (which agree with the nonlinear $\sigma$
model results in
Ref.~\onlinecite{DomRead1989}):
$c_1^{\mbox{\scriptsize LSWT}}=\frac{3\sqrt{3}}{2}J_1S$
and $c_2^{\mbox{\scriptsize LSWT}}=c_3^{\mbox{\scriptsize LSWT}}=
c_1^{\mbox{\scriptsize LSWT}}/ \sqrt{2}$.
For large spin $S$ we have good agreement
with these results. We have also examined the ratio
of the two spin-wave velocities, as shown in
Fig.~\ref{fig:c1c23ratiovsS}.


\section*{Discussion}

To summarize:
We have performed a SBMFT analysis of the spin dynamics of the
antiferromagnetic Heisenberg
model on a triangular lattice, expanding around a state with local
$120^\circ$ order.
Considering that the $120^\circ$ state has been favored as a candidate ground
state in several studies, this is a natural approach.
In order to identify the relevant fields in such an expansion,
we have performed
a spin-rotation and we have also neglected the part of the Hamiltonian
which gives no contribution in the classical limit.

In the limit of infinitely large spin we obtain three massless modes in the
dispersion relation of the spin-waves,
exactly recovering the LSWT results, including the spin wave velocities.
This is an improvement on earlier SBMFT calculations on the triangular
lattice~\cite{YoshMiya1991,LefHed1994}, where the $S \rightarrow \infty$ limit
does not come out correctly.
At finite spin, however, two of the modes acquire a mass. This can be
understood as follows:
The effective Hamiltonian in Eq.~(\ref{eqn:effHam}) is invariant under
rotations in the $x-z$ plane, and the $SU(2)$ symmetry of the original
model is lowered to $U(1)$.
Since the $120^\circ$ state breaks this $U(1)$ symmetry, we should
obtain {\em one} Goldstone mode, which agrees with our results for finite spin.
We do not believe that the massiveness of the other modes are due to enhanced
quantum fluctuations at finite spin:
If the effective Hamiltonian
had preserved the
$SU(2)$ symmetry of the full Hamiltonian we should probably
have recovered the three Goldstone modes as expected. We
are also led to this conclusion by the discussion in
Ref.~\onlinecite{ReplytoCommentonChanColeLark1991} where the analogous problem
of the frustrated square lattice model is considered.

We have already mentioned that the term in the full Hamiltonian that we
neglect in order to obtain the effective Hamiltonian is
also neglected in LSWT. Yet, LSWT predicts three massless modes for arbitrary
spin, even though the LSWT Hamiltonian is not $SU(2)$ invariant. From
Fig.~\ref{fig:M2vsS} we see that for
large spin the mass $M_2$ ($=M_3$) goes approximately as $S^{-1.5}$ and
from Fig.~\ref{fig:ratioc1ratioc23vsS} we see that the spin-wave
velocities goes like the LSWT result, that is, they are proportional to $S$.
Thus, the
energy gap goes like $S^{-1.5}*S^2=S^{0.5}$. Since the energy in LSWT
only retains powers
of $S$ larger than or equal to $1$ and the massgap is approximately
proportional to
$S^{0.5}$, one still obtains massless modes in LSWT. In the
same sense that LSWT is correct for large spin (neglecting terms of
order $S^n$, $n<1$) our results are valid for large spins,
neglecting energy terms of the order of the massgap.

It is interesting to note that
the modes at the corners of the Brillouin zone (picture B) describes
fluctuations in the $y$-component
of the spin, while the center mode describes in-plane
fluctuations~\cite{DomRead1989}.
The first term in Eq.~(\ref{eqn:Spinrot}) is independent of
the $y$-component, which means that fluctuations in this
component are described equally well by the effective
and the full Hamiltonian. This makes the value of the spin-wave
velocity at the corner modes more reliable than that at the center mode,
since this last value directly depends on the
difference between the effective and the full Hamiltonian.

It should be interesting to perform an extended analysis of the spin
dynamics above a state with local $120^\circ$ order. Starting
with the approach of Lefmann and Hedeg\aa rd~\cite{LefHed1994},
one may write the Hamiltonian in terms of a singlet field
(also considered by Yoshioka and Miyazaki~\cite{YoshMiya1991})
and a field representing the ferromagnetic correlations, {\em and
then perform a spin rotation}.
Certain pieces of the two fields then
combines to a constant term and the remaining parts can be written in terms
of four effective fields. Diagonalizing this form of the Hamiltonian using the
SBMFT may possibly yield a dispersion relation which
can take all solutions into account {\em within the same mean-field theory}.
The solution presented in this paper is valid for large spin, as we have
 already
argued, but this
approach may help answering for which magnitudes of the spin the solutions
of Ref.~\onlinecite{YoshMiya1991} and Ref.~\onlinecite{LefHed1994} are
 relevant.
We will return to this subject.

\section*{Acknowledgments}
We thank H. Johannesson, P. Fr\"ojdh and A. Chubukov for discussions
and valuable remarks.


\begin{figure}
\caption{Triangular lattice. The vectors $\protect{\bbox{\alpha}}$ point to
half the number of nearest
neighbors of a site. Sites of the three sublattices are indicated
by squares, crosses and
circles respectively. \label{fig:triangular}}
\end{figure}

\begin{figure}
\caption{The Brillouin zones of the triangular lattice (large) and
of the sublattices (small). The numbers $1$ to $3$ are indicating the parts
of the large Brillouin zone that corresponds to the three different modes
in the small Brillouin zone. Three possible vectors $\protect\bbox{Q}$
(see text) are also shown. \label{fig:Brill}}
\end{figure}

\begin{figure}
\caption{The mass $M_2$ ($=M_3$) of the corner modes
(picture B (see text)) vs. spin $S$.
\label{fig:M2vsS}}
\end{figure}

\begin{figure}
\caption{The dispersion relation for spin a) $S=0.5$ and b) $S=200$ at zero
temperature.
The gap at the "ferromagnetic" point is always zero while the gaps at
the "antiferromagnetic" points only tends to zero for large spin.
We use picture B (see text).
\label{fig:Omega}}
\end{figure}

\begin{figure}
\caption{The ratio of the spin-wave velocities $c_m$ $m=1,2,3$ obtained
by our theory (SBMFT) and by LSWT (or by
Ref.~\protect{\onlinecite{DomRead1989}}) vs. spin $S$.
\label{fig:ratioc1ratioc23vsS}}
\end{figure}

\end{document}